\documentstyle[preprint,prb,aps]{revtex}
\tightenlines
\begin{document}
\draft
\title{ Diamagnetic Interactions in Disordered Suspensions of
Metastable  Superconducting Granules
}
\author{A. Pe\~{n}aranda, C.E. Auguet and L.\ Ram\'{\i}rez-Piscina}
\address{
Departament de F\'{\i}sica Aplicada,\\
Universitat Polit\`{e}cnica de Catalunya,\\
Doctor Mara\~{n}on 44, E-08028 Barcelona, SPAIN.}

\date{\today}

\maketitle
\thispagestyle{empty}

\begin{abstract}
The simulation of the transition sequence of superheated
Type I superconducting granules (SSG) in disordered suspensions when
an external magnetic field is slowly increased from zero has been studied.
Simulation takes into account diamagnetic interactions and the
presence of surface defects.
Results have been obtained for the transition sequence and surface fields distribution covering a wide
range of densities.
These results are compared with previous analytical perturbative
theory, which provides qualitative information on transitions and
surface magnetic fields during transitions, but with a range of
validity apparently limited to extremely dilute samples.
Simulations taking into account the complete diamagnetic
interactions between spheres appear to be a promising tool in
interpreting SSG experiments, in applications such as particle
detectors, and in some fundamental calculations of Solid State
Physics.

KEYWORDS: A. Superconductors. A. Disordered systems. D. Phase
transitions.
\end{abstract}

\newpage

\section{Introduction}

     The study of the electric and magnetic properties of
inhomogeneous or disordered systems has been a subject of
long-standing interest in both basic and applied condensed matter
physics.
Historically, studies of these properties have usually dealt with
the determination of effective parameters (typically dielectric
constants) on a longer spatial-scale than the typical scales of the
inhomogeneities or the disorder \cite{garland}. However, in
superconducting granular materials the situation is more complex
and interesting. These systems, in response to applied fields, can
suffer transitions which depend on
details at the disorder scale.  Moreover, as will become
evident below, local fields change at every transition, which makes
the response of the whole system history-dependent. In
this context, mean or effective properties of these materials are
of limited interest.

Measurements on superconducting granular materials can provide
interesting information from a fundamental point of view. For
example, the measurement of the supercritical fields of disordered
suspensions of superconducting granules has yielded determinations
of the Ginzburg-Landau parameters of the superconducting transition
\cite{smith}. On the other hand, the superheated-to-normal phase
transitions of  Type I superconducting suspensions, induced by
irradiation, have served as the basis for the recent development of
particle detectors \cite{girard}; also the irradiation-induced
supercooled normal-to-superconducting transition of granular arrays
is  being explored \cite{meagher} with respect to dark matter
detection in astroparticle physics.

     In a superconducting granular material, the
transition of each granule is determined by its position in its
phase diagram, which is schematically shown in Fig. 1. Type I
superconductors exhibit hysteresis, so a granule can traverse the
equilibrium curve $B_c$ without transiting as long as it remains in
the metastable region. In Fig. 1 we show the case of an isolated
metastable superconducting granule. It can be seen that the actual
normal to superconducting  transition, whether induced by a field
increase $\Delta B$ or heating  $\Delta T$,
depends critically on its location in the (T,B) phase space referred
to  the superheating field curve $B_{sh}$. Theoretically,
this location is given by the bath temperature and the maximum
field on the surface $B_{max}=(3/2)B_{ext}$. Experimentally, the
location generally varies over a small range  of fields as a result
of surface and volume defects which act as nucleation centers
\cite{feder,hueber}.
In contrast, a disordered suspension exhibits a broader range of
transition field values \cite{drukier,furlan}, which is  generally
due both to defects and to diamagnetic interactions between the
superconductors themselves. This spreading depends directly on the
local fields on the surface of the granules, information which is not
directly accessible in experiments, and can reach typical values
of $20\%$.
Analogous behavior
has been found in superconducting granule detectors, in which
transitions occur by incident energy from particles or radiation.
In this case the uncertainty in the minimum energy necessary for
the transition necessarily hinders the interpretation of
the results of these devices \cite{drukier}.
On the other hand, the long-range nature of the diamagnetic
interaction produces changes in the surface field values for the
entire system  as each transition occurs. In consequence, the
suspension disorder and  its effects change with each transition.
Therefore, it is clear that a complete theoretical analysis of
diamagnetic interactions and their effects on transitions
constitutes a first step in the interpretation of experiments
involving this kind of suspensions.

     There exist numerous theoretical results available
 on magnetic or (mathematically
analogous) electric properties of disordered materials.
 These are mainly for the
effective dielectric constant of composites, involving mean field
calculations, such as the classical mean-field Clausius-Mossotti or
Maxwell-Garnett \cite{garnett} approximation, rigorous bounds
\cite{bergman,milton}, or cluster expansions for suspensions of
polarizable spheres \cite{geige86,felderhof}.
Analogous expansions have been developed for the diamagnetic
properties of superconducting sphere suspensions.
In the regime of dilute suspensions, Geigenm\"{u}ller
\cite{geige88}  constructed a perturbative theory which is not limited to
effective properties,  but which calculates statistics of local
surface fields on the granules  and of the transitions induced by
the external field. This theory is  formally based on a cluster
expansion in such a way that all the quantities  are expanded in
powers of the volume fraction $\rho$ occupied by  the
microgranules.
In practice, the expansion is performed up to first order, which
means that only two-body interactions are considered
\cite{geige88}. Nevertheless, it provides the only analytical
framework in which experiments have been interpreted so far.

In this paper we will study the
magnetic properties of a disordered ensemble of (type I)
superconducting granules immersed in an external magnetic field by numerical simulation. We
specifically address the question of the diamagnetic interactions
between granules, and the process of the successive transitions in
the system induced by an increase of the external field.
Relevant information, which is not experimentally accessible, such as local
surface field values, is obtained during transitions.
Conditions on which these transitions occur (essentially the
external field values) are the crucial point in the applications
mentioned above.
The comparison of the simulation results to the perturbative
calculation shows that the problem in interpreting experiments is
twofold. Firstly, direct application of the perturbative results,
which are valid in the dilute regime in principle, uses linear extrapolations of the experimental
results to the zero-density limit as
parameters of the theory. As we will show, the
validity condition of this procedure ({\em i.e.} that experiments lie in the
proper range of validity of the theory) is hardly fulfilled by
current experimental results. Therefore, quantitative predictions
are expected to fail in all density ranges. Secondly, some
interesting results
arise in measurements performed in high-density
conditions, where diamagnetic interactions are most important and
the perturbative theory does not apply.
This is the case for instance of the experimental determination of
the supercooling branch of the phase diagram and its associated
Ginzburg-Landau parameters. \cite{smith}
In each branch of a hystheresis loop, the last transiting granules
are expected to be the most defect-free ones.
Contrary to the superheating case,
in a supercooling situation
these granules are most affected by diamagnetic
interaction of other superconducting granules.
In view of all these circumstances, both the quantitative determination of the
limits of validity of the perturbative calculation, and the use of
simulation to obtain results valid in the whole density regime are
of prime interest.

     The structure of this paper is the following. The  simulation
procedure is described in detail in Section II. Results obtained
for representative cases, presented in Section III, show the
critical importance of diamagnetic interactions in the
transitions of disordered superconducting suspensions, and
quantitatively demonstrate the limited range of validity  of the
perturbation theory.
In Section IV some conclusions are drawn from this work.

\section{Simulation details}

We performed
simulations of dispersions of $N$
superconducting spheres of radius $a$ sited in positions given by ${\bf
R}_i$ in a thin cylindrical sample of thickness $L$ and radius
$\alpha \times L$. Positions were chosen at random in the prescribed
volume excluding configurations in which the distance between the centers
of any couple of spheres is $d \leq 2a$.
An external field $B_{ext}$ applied
to the system was considered.
The radius of the spheres was considered much greater
than the London penetration length, and the transitions of each sphere
to the normal phase are completed when the local magnetic field on any
point of its surface reaches a threshold value $B_{th}$. Hence, we
do not consider partial transitions to the intermediate state. This
is justified when the local magnetic field applied over a sphere is
greater than the critical magnetic field $B_c$, and thus the sphere
is in a metastable state\cite{geige88}.
Neither the possible effects of diamagnetic contact and associated
percolation phenomena for very concentrated configurations were considered.
The magnetic field ${\bf B}({\bf r})$ can be determined from a
scalar potential $U({\bf r})$ \begin{eqnarray}
{\bf B}({\bf r})= - {\bf \nabla }U({\bf r})
\label{potential},
\end{eqnarray}
which satisfies the Laplace equation
\begin{eqnarray}
\nabla^2 U({\bf r}) = 0.
\label{laplace}
\end{eqnarray}
with the following
boundary conditions. Firstly, for any superconducting  sphere the
magnetic field is tangential to the surface, {\it i.e.} the normal
derivative of the potential vanishes there. Secondly, the value of
the field very far from the sample should match ${\bf B}_{ext}$:
\begin{eqnarray}
U({\bf r})\rightarrow -{\bf r}\cdot {\bf B}_{ext}\;\;(r\rightarrow
\infty ).
\end{eqnarray}
The scalar potential $U({\bf R}_j+{\bf r}_j)$ near sphere $j$ can
be  expanded in multipoles \cite{geige88,geige89}
 which, introducing the boundary conditions at the surface of the
sphere, can be written as
\begin{eqnarray}
U({\bf R}_j+{\bf r}_j)=\sum_{\lambda =1}^\infty \sum_{\mu =-\lambda
}^\lambda Y_{\lambda \mu }(\hat r_j)\ c_{\lambda \mu }(j)\left\{
\left(
\frac a{r_j}\right) ^{\lambda +1}+\frac{\lambda +1}\lambda \left(
\frac{r_j}a%
\right) ^\lambda \right\} +K(j),
\label{expansion}
\end{eqnarray}
where $Y_{\lambda \mu }(\hat r_j)$ are spherical harmonics, and
$c_{\lambda \mu }(j)$ and $K(j)$  are the coefficients of the
expansion. There is one of these expansions for each sphere.

From all the expansions, and employing the boundary conditions, the
coefficients satisfy the following equations \cite{geige89}:
\begin{eqnarray}
K(j)=-{\bf R}_j\cdot {\bf B}_{ext}+\sum_{k\neq j}\sum_{\lambda
=1}^\infty
\sum_{\mu =-\lambda }^\lambda A_{00\,\lambda \mu }(j,k)\ c_{\lambda
\mu }(k)
\label{eq.k}
\end{eqnarray}
\begin{eqnarray}
\frac{\lambda +1}\lambda \ c_{\lambda \mu }(j)=-\sqrt{\frac{4\pi
}3}B_{ext}\
a\ \delta _{\lambda 1}\delta _{\mu 0}+\sum_{k\neq j}\sum_{\lambda
^{\prime
}=1}^\infty \sum_{\mu ^{\prime }=-\lambda ^{\prime }}^{\lambda
^{\prime
}}A_{\lambda \mu \,\lambda ^{\prime }\mu ^{\prime }}(j,k)\
c_{\lambda
^{\prime }\mu ^{\prime }}(k)
\label{eq.c}
\end{eqnarray}
where the constants $A_{\lambda \mu \,\lambda ^{\prime }\mu
^{\prime }}(j,k)$ are given in Ref. \cite{penaranda1}. Without loss
of generality, in Eq. (\ref{eq.c})  we have placed the
external field in the $z$-direction.
After determining the values of the unknowns $c_{\lambda \mu}(j)$
and $K(j)$ for a given configuration, it is possible to calculate
the  surface fields from Eq. (\ref{expansion}).

The constants $K(j)$ only give additive contributions to the
potential and do not affect the magnetic field values, so the
problem is, in principle, to solve the infinite set of linear
equations
(\ref{eq.c}) for the unknown $c$'s. In practice one only takes
into account a limited number of multipolar terms according to
the desired precision. Even then, the number of unknowns is so
large that a direct solution of the Eqs. (\ref{eq.c}) is a
formidable task  for configurations with a representative number
$N$ of spheres.
Instead, we employ the following iterative method
\cite{penaranda1}. Eq. (\ref{eq.c}) can formally be written as a
matrix equation for the vector of unknown ${\bf c}$
\begin{eqnarray}
{\bf c} = {\bf b} + A {\bf c}
\end{eqnarray}
whose solution is
\begin{eqnarray}
{\bf c} = (I - A)^{-1} {\bf b}
\end{eqnarray}
which can be expanded as a power series in $A$,
\begin{eqnarray}
{\bf c} = \left( I + A + A^2 + A^3 + ... \right) {\bf b}.
\end{eqnarray}
The simplest way to numerically perform this expansion is to apply
the iteration
\begin{eqnarray}
{\bf c}_{i+1} = {\bf b} + A {\bf c}_i,
\label{iteration} \\
{\bf c}_{0} = {\bf b}.
\end{eqnarray}
The dependence of the $A$  matrix elements on the distance
between the spheres  guarantees the convergence of this expansion,
which is faster for more dilute systems. The desired precision is
achieved by iteratively applying Eq. (\ref{iteration}) until the
change of the coefficients $c$ is lower than a prescribed value.

The procedure in our simulations is then as follows:
$N$ superconducting spheres are placed at random in the desired
geometry according to the value of the given filling factor
$\rho$ (fraction of volume occupied by the microgranules). The
threshold values $B_{th}$ for every sphere are also assigned by
using a given distribution.
Applying the iterative method described above, the values of the
coefficients $c$ are obtained, which enables determination of the local
values of the magnetic field on the surface of any sphere from
Eqs. (\ref{potential},\ref{expansion}).
The maximum value of the surface field for each sphere is then
calculated by standard routines
of minimization of multivariate functions.
The comparison of these
maximum surface fields with the respective values of $B_{th}$
permits selection of the first superconducting sphere that will
transit to normal under an increase of the applied magnetic field.
Furthermore, the precise value of $B_{ext}$ at which the transition occurs
is monitored. Subsequently, the system becomes one of $N-1$
superconducting spheres.
The long-range nature of the diamagnetic interactions  changes the
surface magnetic field values of the remaining superconducting
spheres on any transition. This leads us to repeat
the same calculation process after each transition until all
spheres have transited.

\section{Results and discussion}

For the simulations we have employed a distribution
of values of $B_{th}$ consistent with experimental results
for tin microspheres diluted in paraffin
\cite{mettout}. A parabolic distribution fit is used in the interval
between $0.8 B_{sh}$ and $B_{sh}$ \cite{geige88,geige89}.
This distribution of $B_{th}$ values was obtained from experimental
data corresponding to systems with small $\rho$ extrapolated to
$\rho=0$. The number of initially superconducting spheres employed
in simulations was $N=250$
for dilute dispersions with $\rho$ values from $0.001$ up to
$0.05$, and $N=150$ for denser systems with $\rho$ up to $0.20$.
 For each case we performed
averages over a number of independent configurations
between $2$ to $7$. The geometrical ratio $\alpha$ was chosen to be
$10$. This large value of $\alpha$ makes finite-size effects important
for the larger volume fractions employed. Indeed for $\rho = 0.20$ a
system of $N = 150$ has a width of only $4.6$ sphere radius, and
hence cannot be considered as truly $3$- dimensional. However, it also
 avoids the appearance at these densities of percolating clusters
associated to diamagnetic contact, which in fact has been ignored in
simulations.

The fraction $f$ of remaining superconducting spheres during an
increase of the external field is shown in Fig. \ref{FBext} versus
$B_{ext}$ for different values of $\rho$.
The furthest to the right continuous line in this figure shows the
expected behavior
for isolated spheres, for which the maximum surface field is equal
to $3/2 B_{ext}$, and therefore can be directly related to the
distribution of $B_{th}$ values. We see that for the most dilute
case ($\rho=0.001$) the $f$ values closely follow that $\rho=0$
limit, except for a few transitions occurring earlier than expected
corresponding to spheres whose distances to the nearest neighbor
are not very large (about $0.10- 0.15$ times the radius value).
Therefore, for such a dilute case,
the observed spread in the transition field values can be
attributed to surface defects.
However, Fig. \ref{FBext} shows transitions for increasingly lower
external fields as the concentrations of the sample are increased.
This enhanced spread in the transition fields is produced by the
diamagnetic interactions between spheres in these more densely
packed configurations, which generate local surface fields much
higher than the externally applied field and are the origin of the
observed dependence on the sample filling factor.
We see that diamagnetic interactions start to be the most important
factor in transition spreading for filling factors of a few per
cent.
Indeed half of the spheres have undergone transitions at
$B_{ext}=0.48 B_{sh}$ for $\rho=0.20$, while for  $\rho=0.001$ a
field $B_{ext}=0.60 B_{sh}$ is required. Similar behavior was observed in the
experimental results of Dubos and Larrea \cite{private}.
In these experiments half of the spheres transited at $B_{ext}=0.48
B_{sh}$ for $\rho=0.25$, at $B_{ext}=0.50 B_{sh}$ for $\rho=0.20$,
and $B_{ext}=0.53 B_{sh}$ for $\rho=0.04$. This agreement confirms
that the mechanisms implemented in the simulations (presence of surface
defects and influence of diamagnetic interactions) are essentially
correct.

     These results suggest an analysis of our results in terms of the
perturbative theory of Geigenm\"{u}ller \cite{geige88}, which takes
into account these mechanisms under the same hypotheses (ignoring
partial transitions, diamagnetic contact, etc) assumed in
our numerical simulations. This theory
 performs a systematic expansion
of quantities such as the surface field distribution and the
transition sequence in powers of the filling factor.
Within the framework of this expansion we can write \cite{geige88}:
\begin{equation}
f(B_{ext},\rho) = f_{0}(B_{ext}) + \rho f_{1}(B_{ext})+ O(\rho^{2})
\label{fexpan}
\end{equation}
where the zeroth order is the same $\rho = 0$ prediction
represented in Fig. \ref{FBext}.
This expansion shows how the distribution of $B_{th}$ can be
obtained from experimental data by performing measurements on
samples of different densities and extrapolating the results to
$\rho \rightarrow 0$ \cite{geige88,geige89,mettout}.
In this expansion, each order can be calculated from
this distribution of threshold fields $B_{th}$, and involves
increasingly higher order contributions both in the number of
spheres and in multipolar interactions. In the present state of the
theory, calculations are done up to first order, which is equivalent
to considering only two-body interactions.

The comparison between our simulations and the results of the
Geigenm\"{u}ller theory should permit one to define the range of
validity of the linear approximation, and provide an insight into the
effects of higher order terms.
To this end Fig. \ref{FBext} also shows the predictions of Eq.
\ref{fexpan} and Ref. \cite{geige88} for $\rho=0$, $0.01$, $0.05$ and
$0.10$ as continuous lines. We see that the theory does contain the
observed trend of transitions to occur for lower external fields
due to diamagnetic interactions, but agreement does not seem to
be quantitative except for extremely dilute samples.
For the $5\%$ case agreement appears to be rather poor and it is
worse for denser systems.

A more appropriate test of the linear approximation involved in Eq.
\ref{fexpan} is the study of the dependence of $f$ on the density
$\rho$ for different values of $B_{ext}$. This is shown in Fig.
\ref{frho}, where symbols represent simulation results
and lines are the perturbative predictions of Eq. \ref{fexpan}.
Note that the evolution of a system during successive transitions is
represented by points at constant $\rho$ and increasing values of
$B_{ext}$. We see that the perturbative calculation of Eq.
\ref{fexpan} provides a correct qualitative picture of the
transitions. However, for intermediate values of $B_{ext}$, the
differences between theory and simulations begin to be
non-negligible at volume fractions between $2$  - $5\%$. These
results further indicate that $\rho \rightarrow 0$ extrapolations
of experimental results, performed in order to obtain information
on the distribution of values of $B_{th}$, should only be made with
very dilute systems. This may explain the apparent discrepancies
between theory and experiment found in Refs.
\cite{geige88,mettout}, where the values of $f_0$
and $f_1$ were evaluated by linear extrapolations of experimental
data up to $\rho = 5\%$. In view of Fig. \ref{frho}, this procedure
should yield erroneous results in the range of the most interesting
values of $B_{ext}$, where transitions mostly occur. It is
precisely in this range where the contribution of $f_1$ in Eq.
\ref{fexpan} is more important, so it is here where a breakdown of
the perturbative scheme is expected \cite{geige88}. We will address
this point below.

     The study of the maximum values of the magnetic field on the
surface of the granules is the key point in analyzing the
transitions,  as pointed out above. This is a quantity that is not
accessible to direct experimental measurements. The theory of
Geigenm\"{u}ller provides estimates of the distribution of values
of such maximum fields in the
small-density expansion \cite{geige88}:
\begin{equation}
P(B_{max},\rho) = P_{0}(B_{max}) + \rho P_{1}(B_{max}) +
O(\rho^{2})
\label{Pexpan}
\end{equation}
In this equation $P(B_{max},\rho)$ stands for the fraction of
superconducting spheres with a maximum surface magnetic field
smaller than $B_{max}$. We have omitted the implicit dependence of
such quantities on $B_{ext}$. $P_{0}(B_{max})$ is the step function
$\theta (B_{max}-3/2B_{ext})$, {\it i.e.} the result for isolated
spheres. The linear term describes the broadening of the
distribution due to magnetic interactions \cite{geige88}. These
interactions change at each transition during the increase of the
external field, and therefore are history dependent. This is the
reason why the broadening described by $P_{1}$ depends on
$B_{ext}$.

The problem of a systematic expansion in $\rho$ of quantities such
as the distribution $P(B_{max},\rho)$ is that by its own nature the
zeroth order is discontinuous, while the effect of a finite $\rho$
should smooth it. However, it is impossible that a linear correction
such as Eq. \ref{Pexpan} could do this job for all values of $\rho$.
In fact, looking closely at the prediction of Eq. \ref{Pexpan} around
$B_{max}=3/2 B_{ext}$ we can check that there exists a discontinuity
and the distribution function ceases to be monotonous. This
breakdown of the perturbative theory is shown in Fig. \ref{criteria},
 where we represent the predictions of the theory
for a dilute case ($\rho=0.01$) and a concentrated case
($\rho=0.20$) at an applied external field $B_{ext}=0.5B_{sh}$. At
this value of $B_{ext}$ all transitions are exclusively due to
diamagnetic interactions, and whereas for dilute systems only a
small fraction have occurred, for more concentrated configurations
the system is deeply immersed in the transition sequence.
We see that two branches appear on both sides of the
discontinuity at $B_{max}=3/2 B_{ext}$. While both branches are
close to each other for small $\rho$ and one could use some
criterion to connect them, for large $\rho$ this is not the case.
We have chosen to use the right-hand side branch in such situations to
extract quantitative information and to compare it with simulations.

In Fig. \ref{PB02} we show simulation and perturbative results of
the distribution $P$ of maximum surface fields for systems at
$B_{ext}=0.2 B_{sh}$ and different initial values of $\rho$. These
results are almost identical to the results without transitions of
Ref. \cite{penaranda1}. For this value of the external field, only
a reduced number of spheres  have already transited, and what one
sees is the result of the diamagnetic interactions on almost
completely disordered configurations of superconducting spheres.
Again, although the perturbative theory qualitatively describes the
diamagnetic effects, namely the broadening of the maximum field
distribution, it only appears as quantitatively correct for very
small values of the occupied volume fraction.
In Fig. \ref{PB06} we show the analogous results for an external
field $B_{ext}=0.6 B_{sh}$.
For this field the system has already suffered a large number of
transitions except for the extremely dilute configurations.
Here the convergence of the perturbative expansion is expected to
be rather limited, and, in fact, its non-analytical behavior at
$B_{max}=3/2 B_{ext}$ is  clearly seen. However the perturbative
theory (the right-hand side branch) continues to provide  satisfactory
qualitative predictions. Note that the maximum field distributions are much narrower after this increase
of the
external field.
In fact, when a large number of spheres have transited induced by
the external field, those that remain superconducting are expected
to form quite ordered configurations \cite{penaranda2,valette}, and,
as a consequence, the surface field values become more uniform in
the sample \cite{penaranda2}. This feature is captured by the
perturbative theory.

In order to show the ordering effect of the
transitions more clearly we present the evolution of the distribution $P$ during
the increase of $B_{ext}$ for three systems with representative
values of the filling factor: $\rho = 0.001$ (Fig. \ref{ro001}),
$\rho = 0.01$ (Fig. \ref{ro01}) and $\rho = 0.20$ (Fig.
\ref{ro020}).
For each filling factor, both simulations and perturbative results
are shown for values of the external field ranging from $0.2
B_{sh}$ up to  $0.6 B_{sh}$, for which  most transitions occur.
We see again that the broadening of the field distribution is
drastically reduced as the external field is increased. The results
indicate discrepancies between the perturbative theory and the
simulations  for the lowest values of $B_{ext}$ even for very
dilute systems.  However, agreement improves as
$B_{ext}$ increases and the systems undergo transitions.

The results obtained for the most dense system with a volume
fraction $\rho=0.20$ (Fig. \ref{ro020}) are particularly
interesting: the large discrepancies observed at $B_{ext}=0.2
B_{sh}$ evolve to a fair agreement for $B_{ext}=0.5 B_{sh}$ or
$B_{ext}=0.6 B_{sh}$. In Fig. \ref{bmeanbex} we characterize the
distribution of maximum surface fields for this filling factor by
its mean and its standard deviation values, which are represented versus
the increasing external field and compared with the perturbative
theory. One feature of the theory that is visible here is that it
predicts transitions only for external fields greater than $0.2
B_{sh}$. We see that the surface fields approach the isolated
sphere values as $B_{ext}$ increases,
whereas the standard deviation approaches zero indicating an
increase of the uniformity of the system. We also see that the
theory approaches the simulation results quite satisfactorily for
external fields greater than $0.4 B_{sh}$.

 This agreement after transitions between simulations and a small
density approximation is quite surprising for such a dense case,
and appears to be better than that obtained for low densities and low
external field. One possible explanation could be that when a large
number of transitions have occurred, the effective filling factor
(if we only take into account the remaining superconducting
spheres)
is smaller, and one could expect a better convergence of the
$\rho$-expansion.
However, this agreement corresponds to systems with effective
volume fractions after  transitions that are higher than in the low field
systems represented in  Figs. \ref{ro001} and \ref{ro01} for
$\rho=0.001$ and $0.01$. This can be seen in Fig. \ref{estadis},
where we have represented the mean and standard deviation of the
maximum surface fields versus the effective filling factor
$\rho_{ef}$ for the dense system with initial $\rho=0.20$. In this
figure the evolution of the system is to decreasing values of
$\rho_{ef}$. We already see good agreement for $\rho_{ef}=0.1$,
which is a rather large value. To show this, in Fig. \ref{ro020}
we have also represented the distribution of maximum fields for a
configuration with the same $\rho_{ef}$ (slightly lower than $0.1$)
than that of  $B_{ext}=0.5 B_{sh}$, but with positions completely
at random, and the corresponding predictions from perturbative
theory.
Note the good agreement between theory and simulations for the
system which  has been ordered by transitions compared with the
discrepancies observed for the last random configuration with the
same effective filling factor.
We can conclude that
the perturbative theory implicitly includes the observed tendency
to form ordered configurations of remaining superconducting
spheres, and that it describes these resulting
configurations quite well, even  if they arise from much more concentrated
initial systems. However this good behavior of the theory is not
completely explained by the decrease of the effective filling
factor during the transition process.

\section{Summary and Conclusions}

     In summary, we have performed numerical simulations of
disordered suspensions of superheated superconducting granules
(SSG), transiting to normal when an external magnetic field is
slowly increased from zero. Transitions are controlled by the local
surface magnetic fields, which depend in a non-trivial way on the
geometrical configuration of the superconducting granules via
strong diamagnetic interactions, and the presence of surface
defects. Performed simulations employ complete resolutions of the
Laplace equation for the magnetic field with suitable boundary
conditions on the surface of the superconducting spheres, and
therefore are expected to provide results which are not limited to dilute
samples. Consistent comparison with previous experimental results
indicates that the simulations capture the mechanisms involved in
the real system, and their use in simulations appears to
be essentially correct. A better comparison with experimental results
might need a revision of the $B_{th}$ distribution.
Also the employ of a larger number of spheres in simulations could be
necessary to reduce finite-size effects \cite{penaranda1}.
On the other hand, the numerical method can be straightforwardly
generalized to
SSG configurations with spheres of different sizes.
Preliminary simulations with realistic size distributions
show
the same qualitative behaviour than obtained with spheres
of equal radius.

Transition sequences as a function of the
(increasing) applied field have been obtained for a large range of granule
concentrations. Distributions of local maximum surface fields, a
relevant quantity in interpreting SSG experiments but not directly
measurable, have also been obtained.
Results indicate a non-negligible effect of diamagnetic
interactions between spheres, even for dilute systems, manifested in
transitions occurring for lower external fields. These interactions
appear as the most dominant factor in transitions for filling
factors starting at a few per cent.
The distribution of maximum surface fields becomes wider for denser
configurations owing to diamagnetic interactions, but successive
transitions induced for these interactions reduce the dispersion
of these surface fields. This homogenising effect is associated with
a positional ordering of the remaining superconducting spheres, and
should be an important factor in reducing the energy uncertainty in
detection applications of SSG systems.

We have compared simulation results of both transitions and surface
field distributions with the analytical cluster expansion of
Geigenm\"{u}ller \cite{geige88}. Theory and simulations share the
same mechanisms (diamagnetic interactions and surface defects), and the
same hypothesis (not considering partial transitions nor diamagnetic
contacts) in modeling SSG systems. However the theory has been developed
up to first order in the filling factor, which limits the calculation to
two body interactions.
Simulation results show that the perturbative theory qualitatively
predicts the behavior of the system, namely the dependence of the
transitions and the maximum surface fields on the concentration of
the sample and on the value of the external applied field.
However, perturbative theory is quantitatively correct only for
very dilute samples, with occupied volume fractions of at most
$1$ - $2\%$. This range of validity increases for systems initially
at higher densities after having undergone a large number of
transitions. The ordering of these resulting configurations is well
described by the theory.

As a final conclusion, diamagnetic interactions appear to be a
factor of fundamental concern in SSG systems. Results from
perturbative expansions, although providing a very useful framework
for analyzing experiments, have a rather limited range of validity,
and should be used with caution in obtaining quantitative
information. In particular, extrapolation to the zero concentration
limit of experimental data should only be performed for very dilute
samples, within the validity range of the expansion.
The use of simulations to obtain results for all densities should
become an essential tool in interpreting SSG experiments. In this sense,
generalization of the theoretical framework in which both simulations
and perturbative theory are based could be essential to analyze very
concentrated systems, in which partial transitions due to diamagnetic
contact can occur.

These results, or more properly the numerical techniques in
obtaining them, may also be of interest in a re-examination of the
earlier Ginzburg-Landau  parameter determinations from
multi-grain measurements  of the supercooled-to-superconducting
transition. This numerical approach should  also be of use in areas
of condensed matter physics other than  superconductivity, such as
viscous fluid flow through a disordered porous medium,
where similar equations arise, or investigation of the magnetic
properties of ultrathin ferromagnetic films \cite{hu}.

\acknowledgments
We would like to thank T. Girard for helpful discussions and manuscript
corrections. We acknowledge financial support from Direcci\'on
General de  Investigaci\'on Cient\'{\i}fica y T\'ecnica (Spain)
(Projects BFM2000-0624-C03-02 and PB98-1203),
Comissionat per a Universitats i Recerca (Spain)
(Projects 1999SGR00145 and 2000XT-0005),
and European Commision (Project TMR-ERBFMRXCT96-0085). We also
acknowledge computing support from Fundaci\'o Catalana per a la
Recerca-Centre de Supercomputaci\'o de Catalunya (Spain).

%FIGURE CAPTIONS

\begin{figure}
\caption{
Phase diagram for a type I superconductor. $\Delta B$ and
$\Delta T$ represent the increase of either the magnetic field or
the temperature needed for a transition to the normal state. }
\label{phasediagram}
\end{figure}

\begin{figure}
\caption{
Fraction $f$ of spheres that remain superconducting versus
$B_{ext}/B_{sh}$, after an increase of the external magnetic field
from zero, for different occupied volume fractions $\rho$. Symbols
represent simulation results.
From right to left, continuous lines correspond to predictions of
Eq. \ref{fexpan} and Ref. \cite{geige88}
%Eq. 6 and Ref. [15].
for $\rho=0$, $0.01$, $0.05$ and $0.10$. The case of $\rho=0$ is the dilute
limit, {\it i.e.} assuming a maximum surface field of $1.5 B_{ext}$
for  all the spheres.
}
\label{FBext}
\end{figure}

\begin{figure}
\caption{
Fraction $f$ of spheres that remain superconducting in
function of $\rho$ for several values of the increasing $B_{ext}$. Symbols are
simulation results: ($\circ$) $B_{ext}=0.30 B_{sh}$,
($\times$) $B_{ext}=0.40 B_{sh}$,
squares $B_{ext}=0.50 B_{sh}$,
($\ast$) $B_{ext}=0.53 B_{sh}$,
($\diamond$) $B_{ext}=0.55 B_{sh}$,
($+$) $B_{ext}=0.58 B_{sh}$,
($\triangle$) $B_{ext}=0.60 B_{sh}$,
($\bullet$) $B_{ext}=0.62 B_{sh}$,
full squares $B_{ext}=0.65 B_{sh}$.
From the top to the bottom, continuous lines are predictions of
Eq. \ref{fexpan} and Ref. \cite{geige88}
%Eq. 12 and Ref. [15].
for the same values of $B_{ext}$.
}
\label{frho}
\end{figure}

\begin{figure}
\caption{Fraction $P$ of spheres with maximum surface field lower
than the $x$-axis value (in units of $B_{ext}$), obtained from
Eq. \ref{Pexpan} and Ref. \cite{geige88}, when the external
magnetic field has reached the value $0.5 B_{sh}$, for $\rho=0.01$ and
$\rho=0.20$. The right branch was chosen to extract quantitative
information.
}
\label{criteria}
\end{figure}

\begin{figure}
\caption{Fraction $P$ of spheres with maximum surface field lower
than the $x$-axis value (in units of $B_{ext}$), when the external
magnetic field has reached the value $0.2 B_{sh}$, for several
values of $\rho$.
Continuous lines are the corresponding predictions of
Eq. \ref{Pexpan} and Ref. \cite{geige88}
%Eq. 13 and Ref. [15].
}
\label{PB02}
\end{figure}

\begin{figure}
\caption{Fraction $P$ of spheres with maximum surface field lower
than the $x$-axis value (in units of $B_{ext}$), when the external
magnetic field has reached the value $0.6 B_{sh}$, for several
values of $\rho$.
Continuous lines are the corresponding predictions of
Eq. \ref{Pexpan} and Ref. \cite{geige88}
%Eq. 13 and Ref. [15].
}
\label{PB06}
\end{figure}

\begin{figure}
\caption{Fraction $P$ of spheres with maximum surface field lower
than the $x$-axis value (in units of $B_{ext}$), corresponding to
an initial $\rho=0.001$ for several values of external magnetic
field
Continuous lines are the corresponding predictions of
Eq. \ref{Pexpan} and Ref. \cite{geige88}
%Eq. 13 and Ref. [15]
}
\label{ro001}
\end{figure}

\begin{figure}
\caption{Fraction $P$ of spheres with maximum surface field lower
than the $x$-axis value (in units of $B_{ext}$), corresponding to
an initial $\rho=0.01$ for several values of external magnetic
field
Continuous lines are the corresponding predictions of
Eq. \ref{Pexpan} and Ref. \cite{geige88}
%Eq. 13 and Ref. [15].
}
\label{ro01}
\end{figure}

\begin{figure}
\caption{Fraction $P$ of spheres with maximum surface field lower
than the $x$-axis value (in units of $B_{ext}$), corresponding to
an initial $\rho=0.20$ for several values of external magnetic
field.
The results for a system with the same
number  of spheres that remain superconconducting in the
case of  $B_{ext}=0.5 B_{sh}$ but placed at random are also represented.
Continuous lines are the corresponding predictions of
Eq. \ref{Pexpan} and Ref. \cite{geige88}
%Eq.  13 and Ref. [15].
 Dashed line corresponds to the prediction for the
random configuration.
}
\label{ro020}
\end{figure}

\begin{figure}
\caption{Mean and standard deviation values(in units of $B_{ext}$)
 corresponding to the
distribution of maximum surface fields versus the increasing external
field for initial $\rho=0.20$. The results from the simulations and from the
perturbative theory (GT) are represented.
}
\label{bmeanbex}
\end{figure}

\begin{figure}
\caption{Mean and standard deviation values(in units of $B_{ext}$)
 corresponding to the
distribution of maximum surface fields versus the resulting effective filling
factor $\rho_{ef}$.
The results from the simulations and from the
perturbative theory (GT) are represented
 for a system with initial
$\rho=0.20$ through progressive increasing of the
external field.
}
\label{estadis}
\end{figure}

\end{document}